\documentclass[12pt,preprint]{aastex}
\usepackage{amsmath}

\begin{document}

\title{Gas Accretion by Globular Clusters and Nucleated Dwarf Galaxies and
the Formation of the Arches and Quintuplet Clusters.}

\author{Douglas N. C. Lin}
\affil{University of California,
Lick Observatory, Santa Cruz, CA  95064}
\author{\& Stephen D. Murray}
\affil{University of California,
Lawrence Livermore National Laboratory, P.O. Box 808, Livermore, CA 94550}


\begin{abstract}
We consider here the collective accretion of gas by globular clusters and
dwarf galaxies moving through the interstellar medium.  In the limit of high
velocity and/or sound speed of the ISM, the collective potential of the cluster
is insufficient to accrete significant amounts of gas, and stars within the
systems accrete gas individually.  We show, however, that when the sound speed
or the relative velocity of the ambient medium is less than the central
velocity dispersion of the cluster, it is accreted into the collective
potential of the cluster prior to being accreted onto the individual
stars within the cluster.  The collective rate is strongly enhanced relative
to the individual rates.  This effect may potentially modify the white dwarf
cooling sequence in globular clusters with low-inclination and low-eccentricity
Galactic orbits, and lead to the rejuvenation of some marginally surviving
cores of globular clusters and nucleated dwarf galaxies near the Galactic
center.  Such effects will only occur rarely, but may explain the existence of
clusters of young, massive stars near the Galactic center.  

\end{abstract}

\section{Introduction}
\label{sec:intro}
Stars within individual open and globular clusters are coeval and have
remarkably homogeneous chemical composition (see, eg. Penny \& Dickens 1986;
Richer \& Fahlman 1986; Stetson \& Harris 1988; Ferraro et al.  1991, 1992;
Kraft et al. 1992; Langer et al. 1992; Suntzeff 1993; Quillen 2002;
Wilden et al. 2002).  These properties provide strong constraints
upon theories of star formation and evolution within these objects.
In the most metal deficient globular clusters such as M92, the total
heavy element content is comparable to the output of a small
number of supernova events.  The observed upper limit on the color
spread among their giant branch stars implies their metallicity
dispersion $ \Delta Z <10-20\%$.  Such small spreads place
limits on the accretion of enriched interstellar medium by the main
sequence progenitors, which can be strengthened further using differential
spectroscopy.

The accretion of the interstellar medium has more dramatic effects on
both white dwarf and neutron stars.  The white dwarf cooling sequences
in both open and globular clusters are used as probes to determine the
ages of these oldest stellar subsystems in the Milky Way galaxy (Richer
et al. 1997, 1998; von Hippel \& Gilmore 2000; Kalirai et al. 2001;
Andreuzzi et al. 2002; Hansen et al. 2002).  The intrinsic
luminosity of the white dwarf is, however, typically
$< 10^{-3} L_\odot$.  Accretion of ambient gas at a rate $> 10^{-14}
M_\odot$ yr$^{-1}$ would generate sufficient energy to modify both the
color and magnitude of these degenerate dwarfs.  Although the
expected accretion rate of individual halo stars is several magnitudes
below this value, it is not clear whether gas can be accreted first,
at a sufficiently high rate, into the collective gravitational
potential of individual clusters.  The increased ambient density within
the cluster would then enhance the accretion of the individual stars.

The accretion of ambient gas by moving bodies is a classical problem.
Many studies have been focused on the flow around compact stars with a
point mass potential (cf Frank, King \& Raine 2002). Although clusters
have much larger masses than individual stars, their potential is
relatively shallow.  In this paper we consider the efficiency of
accretion in these cluster potentials.  We are interested in
determining the accretion rate for clusters with a range of velocities
relative to the ambient medium.  Cases with high relative velocity and
sound speed represent globular clusters belonging to the halo and
thick disk populations.  Cases with low relative velocity and sound
speed are more relevant to young open clusters, globular clusters that
have sedimented into the galactic disk, and the residual dense cores
of globular clusters or nucleated dwarf galaxies near the Galactic
center.

In \S2, we briefly recapitulate the relevant equations which describe
the accretion flow.  We deduce analytically a condition for the
clusters to accrete gas collectively.  Because the flow is multi-dimensional
in nature, we adopt a numerical approach to verify our
analytic approximation.  We describe the numerical scheme and the range
of model parameters and the results of these calculations in \S3.
These results are applied, in \S4, to study the accretion of gas onto
sedimented globular clusters as well as residual dense cores of
clusters and nucleated dwarf galaxies near the Galactic center.  In
\S5, we summarize our results and discuss their implications.

\section{Accretion Onto a Shallow Potential}
\label{sec:analytic}
The basic governing equations for the ambient gas are the continuity
and momentum equations.  In order to highlight the dominant physical
effects, we neglect the energy equation and adopt a polytropic
equation of state.  In general, the accretion flow onto a gravitating
body is known to be unstable and unsteady.  Because we are interested
in the time-averaged accretion rates for illustration purpose, we seek
steady state solutions of the governing equations such that
\begin{equation}
\nabla\cdot \rho {\bf u} =0
\label{eq:cont}
\end{equation}
\begin{equation}
\rho {\bf u} \cdot \nabla {\bf u} = - \nabla P - \rho \nabla \phi
\label{eq:momn}
\end{equation}
where $\rho$, $P$, and ${\bf u}$ are the density, pressure, and velocity
of the background gas.  The gravitational potential of the cluster can
be approximated with a Plummer potential,
\begin{equation}
\phi(R)={{-GM_d}\over{\left(R_c^2+R^2\right)^{1/2}}},
\label{eq:phiplu}
\end{equation}
where $M_d$ is the total gravitating mass, $R_c$ is the core radius,
inside of which the density of the gravitating matter is approximately
constant, and $R$ is the distance from the center of the potential.
The Plummer potential is relatively centrally concentrated, and has
been used to model globular clusters \citep{Plummer11}.

Equations (\ref{eq:cont}) and (\ref{eq:momn}) are multi-dimensional in
nature, and they are treated numerically in the next section.  We can,
however, obtain some approximate analytic solutions to delineate
various regimes of interest.  With a polytropic equation of state, in
which $P = K \rho^\gamma$ where $K$ and $\gamma$ are the adiabatic
constant and the polytropic power index, we find from equations
(\ref{eq:momn}) and (\ref{eq:phiplu}) that
\begin{equation}
{\bf \tilde u} \cdot \nabla {\bf \tilde u} = - \tilde c_s^2 \nabla {\rm ln}\rho
- \xi (1 + \xi^2)^{-3/2} \nabla \xi
\label{eq:vnorm}
\end{equation}
where 
\begin{equation}
{\xi=R/R_c}
\label{eq:radius}
\end{equation}
is a dimensionless radius.  The quantities ${\bf \tilde u}$ and $\tilde c_s$
are dimensionless speeds, defined as
\begin{equation}
{\bf \tilde u}={{\bf u}\over{(G M_d / R_c)^{1/2}}},
\end{equation}
and
\begin{equation}
\tilde c_s={{C_s}\over{(G M_d / R_c)^{1/2}}},
\end{equation}
where $C_s$ is the sound speed of the fluid.  From the continuity
equation~(\ref{eq:cont}),
\begin{equation}
{\bf \tilde u} \cdot {\rm ln} \rho = \nabla {\bf \tilde u}.
\label{eq:rnorm}
\end{equation}
We now consider three limiting cases. We first consider the
possibility of very small $R_c$ such that both
$\tilde u(\xi=\infty)\equiv \tilde u_\infty \ll 1$ and
$\tilde c_s (\xi=\infty) \equiv \tilde c_\infty \ll 1$.  In this limit, the
potential essentially
reduces to that of a point mass and the accretion rate is given by
Bondi solution (Bondi 1952; Hoyle \& Lyttleton 1941; Shu 1992), such that
\begin{equation}
\dot M_c = A (\gamma) \pi
\rho(\infty) (C_\infty ^2 + u_\infty^2)^{1/2} R_B^2
\label{eq:mdotc}
\end{equation}
where $A$ is a constant of order unity, $C_\infty$ and $u_\infty$
are, respectively, the non-normalized sound speed and flow velocity.
The Bondi radius is given by
\begin{equation}
R_B={{GM_d}\over{\left(C_\infty^2 + u_\infty^2 \right)}},
\label{eq:bondi}
\end{equation}
so that
\begin{equation}
\dot M_c = A (\gamma) 10^{25} \left( {n_\infty \over 10^2 {\ \rm cm}^{-3}}
\right) \left({ M_d \over 10^6 M_\odot} \right)^2
\left( 1 + {C_\infty^2 \over u_\infty^2} \right)^{-3/2}
\left( {u_\infty \over 10 {\ \rm km\ s}^{-1}} \right)^{-3} {\rm g\ s}^{-1}
\label{eq:mdotc1}
\end{equation}
where $n_\infty = \rho_\infty/m_h$ and $m_h$ is the mass of the
hydrogen atom.

We note that the above form of $R_B$ is derived for a point-mass
potential, and so is only a good approximation for our models when
$R_B$ significantly exceeds $R_c$.  We now consider two limiting cases
with finite values of $R_c$. In case 2, we consider
$\tilde c_s (\infty) > > \tilde u(\infty)$, such that the flow is nearly
spherically
symmetric and equations(\ref{eq:vnorm}) and (\ref{eq:rnorm}) reduce to
\begin{equation}
{d {\rm ln} \tilde u \over d {\rm ln} \xi}
= \left( {1 \over \tilde u^2 - \tilde c_s ^2} \right)
\left( 2 \tilde c_s^2 - {\xi^2 \over (1 + \xi^2) ^{3/2}} \right).
\label{eq:dlnvdr}
\end{equation}
The maximum value of $\xi^2/(1 + \xi^2)^{3/2}$ occurs at $ \xi= \xi_m \equiv
{\sqrt 2}$.  Because $\tilde c_s $ is a monotonically decreasing function of
$\xi$, the numerator of the right side of equation~(\ref{eq:dlnvdr}) remains
positive definite provided
\begin{equation}
C_s (\infty) > 3^{-3/4} (G M_d/ R_c)^{1/2}.
\end{equation}
Consequently, $\vert \tilde u \vert < \tilde c_s$ throughout the flow.  The
ambient gas therefore collects into a quasi-static atmosphere, with a density
distribution given by
\begin{equation}
\rho (R) = \left[ \rho_\infty ^{\gamma -1 } + \left( {\gamma-1 \over K
\gamma} \right) {G M_d \over (R^2 + R_c^2)^{1/2}} \right]^{1 /
(\gamma-1)} = \rho_\infty \left[ 1 + {\gamma-1 \over \tilde c_s^2 (\infty)
(1 + \xi^2)^{1/2}} \right] ^{1/(\gamma-1)}
\end{equation}
where $\rho_\infty$ is the density of the ambient medium.  If
$C_\infty = C_s (\infty) > (G M_d / R_c)^{1/2}$, the density
enhancement throughout the cluster is very small.

The relative motion of typical globular clusters with respect to the
interstellar medium is generally supersonic, {\it i.e.} $\tilde c_\infty
\ll u_\infty$.  The flow pattern for passage around the cluster with
a large relative velocity, $u_\infty$, is multi-dimensional and
complex.  In line with the conventional treatment of Bondi-Hoyle
accretion flow, it is intuitively tempting to modify the transition
condition to be
\begin{equation}
C_\infty^2 + u_\infty^2 > G M_d / R_c.
\label{eq:bcond}
\end{equation}
When this condition is satisfied, it is more appropriate to use the
Bondi-Hoyle accretion formula for individual stars
\begin{equation}
\begin{array}{cl}
\dot M_\ast &
= B (\gamma) \pi \rho (R) (G M_\ast)^2 (C_\infty^2 + u_\infty^2)^{-3/2} \\
&
=10^{11}B(\gamma) \left( {n_\infty \over 10^2 {\ \rm cm}^{-3}} \right)
\left({ M_\ast \over M_\odot} \right)^2
\left( 1 + {C_\infty^2 \over u_\infty^2} \right)^{-3/2}
\left({u_\infty \over 10 {\ \rm km\ s}^{-1}}\right)^{-3} {\ \rm g\ s}^{-1}
\end{array}
\label{eq:mdots}
\end{equation}
where $B (\gamma)$ is a constant of order unity (it is not the same as
$A$ due to the difference in the potential) and $M_\ast$ is the mass
of the star.  The total accretion rate of the cluster is
\begin{equation}
\dot M_t = N \dot M_\ast
\label{eq:mdotsum}
\end{equation}
where the number of stars in the cluster is $N \sim M_d/M_\ast$.  But
when equation~(\ref{eq:bcond}) is not satisfied, $\dot M_c$ in
equation~(\ref{eq:mdotc}) is the more appropriate rate for the cluster.
Note that
\begin{equation}
\dot M_c \sim N \dot M_t > > \dot M_t.
\end{equation}
This inequality clearly demonstrate that the collective effect of the
cluster would be greater than sum of the individual stellar
contributions if the it moves subsonically through the interstellar
medium.

In the next section, we will carry out 3-D numerical computations to
verify the condition in equation~(\ref{eq:bcond}).  Here we provide some
analytic approximation for the limit that $u_\infty > > C_\infty$.
Without the loss of generality, we consider a case 3 in which the
ambient gas is approaching a cluster in the x direction and the downstream
flow velocities can be written as $u_x = u_\infty + \delta u_x$
and $u_y = \delta u_y$.  Neglecting contribution from the pressure term in
equation~(\ref{eq:vnorm}), we find that along the streamline with an initial
impact distant $R_i$,
\begin{equation}
{\delta u_x \over u_\infty} \simeq - {G M_d \over u_\infty^2 (R_i^2 +
R_c^2)^{1/2}}.
\end{equation}
In the absence of shock dissipation, the potential vorticity,
$\omega/\rho$, and the Bernoulli energy, $\vert u \vert^2/2 + W +\phi$,
are constant of motion along stream lines.  In the limit that the
enthalpy $W = \int dP/\rho$ is small compared with the kinetic and
potential energy, the Bernoulli constant implies that
$\vert \delta u_y / u_\infty \vert \sim \vert \delta u_x / u_\infty\vert$.
The deflection angle of the stream line $\delta y_u/ u_\infty$ is small in
the limit $u_\infty ^2 > G M_d / R_c$.  We also note, from the
conservation of the potential vorticity, that, in the high velocity
limit, the modification of $\rho$ is small despite the slight
convergence of the streamlines.

\section{Numerical Models}
\label{sec:NumModels}

\subsection{Method}
The above analytic approximation is greatly simplified.  The flow is
actually multidimensional in nature.  We therefore proceed by considering
multidimensional simulations of accretion by a defined potential moving
relative to a dense gas cloud.  The results found here shall be used to
validate the analytic estimates of \S~2.

The numerical code we use is Cosmos.  It is a massively parallel,
multidimensional, radiation-chemo-hydrodynamics finite difference code
developed at Lawrence Livermore National Laboratory.  This scheme, and tests
of the code are described in \cite{AF03}, and \cite{AFM03}.  It has also been
used to study the evolution of supernova-enriched material in dwarf
galaxies by \cite{FMAL03}, and to study Roche Lobe overflow from
cluster and galactic potentials by \cite{MDL03}.  The reader is
referred to those papers for more details on Cosmos.  We discuss here
the settings used for the current work.

For a spherically symmetric potential, the unperturbed flow would be
expected to be two-dimensional.  In flows where $u_\infty \sim C_\infty$,
however, an axisymmetric flow pattern may be unstable
(Landau \& Lifshitz 1959; Batchelor 1967).  Our objective is to determine the
time averaged accretion rate rather than to identify any flow instability.
Because Cosmos is written in Cartesian coordinates, however, we cannot
impose cylindrical symmetry upon the problem, and so the simulations are
carried out in three-dimensions.  So as to reduce computational expense, the
models are quadrants, with one-fourth of the regions included.

We wish to examine the ability of a potential to accrete gas as
a function of the relative speed of the potential through the gas, and
the gas temperature.  To improve the controlled nature of the models,
we do not include radiative heating or cooling.  The gas, instead,
evolves adiabatically.  The effects of radiative equilibrium are
approximated by having the gas evolve with an adiabatic constant
$\gamma=1.01$, giving approximately isothermal behavior.
In cases where sufficient gas is
accreted for it to become self-shielded, cooling could decrease the
temperature of the gas significantly, potentially enhancing the
accretion rate beyond the values computed here.

Rather than examine the motion of a potential through a static cloud, we
take the simpler, equivalent approach of viewing the encounter within the
frame of the cluster, in which a dense cloud of gas sweeps over the fixed
potential.  The model clouds are extremely simple.  They are slabs, with
thickness $l=100$~pc, moving across the problem with speeds of
$u_\infty$.  

As discussed in \S~2, the important quantities are the relative speed of the
cloud and the potential, $u_\infty$, and the sound speed of the gas,
$C_s$, and these quantities are varied between the models.  The gas within
the dense slab has temperature selected to give the desired value of
$C_s$ and a density chosen to cause it to be in pressure equilibrium with
a hot background gas.  The hot background gas in the problem has
$T=10^6$~K and $\rho=1.6\times10^{-25}$~g~cm$^{-3}$.  The pressure of the gas,
$nT=10^5$, is chosen to be comparable to that expected within the central
regions of galaxies, or in dense cooling flows.

As stated above, accretion by a potential moving relative to a gas cloud is
complex.  In the frame of the potential, the gas streamlines are bent towards
the cluster center.  Some shall intersect the center, while others converge
along a line behind it.  The convergence speed of the gas determines the
reduction in its velocity relative to the potential due to shocks, and
therefore whether or not the gas is accreted.  In order to ensure that we
are accurately measuring the mass of accreted gas, we carry out our models
until the dense slab of gas has completely swept past the potential.
The increase in the amount of gas within the potential radius as compared
to that present initially, $M_{acc}$, is computed, and listed for each of 
our models. If the value of $M_{acc}$ is seen to be changing at the end of
the simulation, the model is run for additional time.  The minimum time for
the simulations is a few times $l/u_\infty\approx 100{\rm Myr}\left(
{{u_\infty}\over 1{\rm km/s}}\right)$.

For the models examined here, the total gravitating mass,
$M_d=7\times10^5$~M$_\odot$, and the core radius of the potential,
$R_c=10$~pc.  To approximate the effects of external gravitational
fields, the potential is flattened (the gravitational force goes to
zero) beyond an assumed tidal radius $R_t=80$~pc.  The central
velocity dispersion $\sigma = (GM_d/R_c)^{1/2} =16$~km~s$^{-1}$, similar
to the values within many globular clusters (Pryor \& Meylan 1993;
Djorgovski 1993).  In some centrally condensed clusters such as M15,
$\sigma$ can reach $20$~km~s$^{-1}$.  In some nucleated dwarf
galaxies, $\sigma>30$~km~s$^{-1}$ (Peterson \& Caldwell 1993; Geha,
Guhathakurta, \& van der Marel 2002).  In contrast, $\sigma$ in
the Pleiades and Hyades clusters are $0.6$ and $0.2-0.4$~km~s$^{-1}$,
respectively (Perryman et al. 1998; Chen \& Zhao 1999; Madsen, Dravins,
\& Lindegren 2002).  Despite these differences, the discussion in
the previous section indicate that the governing equations can be
normalized with dimensionless parameters and the results obtained
therefrom can be scaled to the appropriate limits.

Self-gravity of the gas is not included in the models.  This approximation
should be adequate for most of our models, for which the accreted mass is
less than the mass responsible for the potential.  For two of the models
discussed below,
the accreted mass exceeds that of the potential.  For those models, the
inclusion of self-gravity would further enhance the
accretion rate.  Both because of this, and the additional cooling
discussed above, the accreted masses for those two models should be
treated as lower limits.

The physical dimensions of the models are 400x200x200~pc, with resolutions
of 2~pc.  The core radius, $R_c$, is therefore resolved by
only 5 zones.  While not highly resolved, the fraction of the accreted mass
contained within the core is generally small, and so the limited resolution
within the cores does not affect our conclusions.

\subsection{Model Results}
\label{sec:results}

The parameters of the models which we have run are shown in
Table~\ref{tab:results}.  In
the table are listed, for each model, the speed of the gas relative to
the potential, $u_\infty$, the sound speed of the gas, $C_s$, the
Bondi radius (as defined in equation~\ref{eq:bondi}), $R_B$, the crossing
time of the potential across the dense slab at $u_\infty$,
$\tau_c\equiv l/u_\infty$, the sound crossing time across $R_B$,
$\tau_s\equiv R_B/C_s$, the amount of gas predicted by
approximate theory to be accreted (see below), $M_p$, and the actual amount
of gas accreted in the model, $M_{acc}$.  The value of $M_{acc}$ is taken as 
the increase in mass contained within $R_B$ relative to that present at the
beginning of the simulation, and we ensure that the value is not changing
by the end of the simulation (see above).

The parameters of the models are chosen so as to span the range of possible
behaviors.  In Model~1, $u_\infty>\sigma$, while in Model~6, $C_s>\sigma$.  In
both models, little accretion is expected, but in Model~1 this is because of
the high relative velocity, while in Model~6 it is due to the high sound speed
of the gas.  In Model~2, $u_\infty\approx\sigma>>C_s$, while in Model~5,
$C_s\approx\sigma>>u_\infty$.  Both models represent marginal conditions for
accretion, but again for different reasons.  In Model~3,
$u_\infty=C_s\ll\sigma$, and so that model is expected to represent an ideal
situation for substantial
accretion.  In Model~4, $u_\infty=C_s\approx\sigma$, and so, again, that model
represents a marginal situation for accretion.

Snapshots showing the evolution of Model~2 are shown in
Figure~\ref{fig:model2rho}.  In the figure, the dense slab of gas can be
seen sweeping over the gravitational potential, which is visible due to its
effect upon the density of the gas moving past it.  Because
$u_\infty\approx\phi^{1\over2}$, some gas is pulled ahead of the cloud by
the potential, and a large fraction of the gas within the tidal radius of the
potential is accreted during the cloud passage.  Once the dense slab of
gas is past the potential, a small, dense core of accreted gas remains in
the center of the potential.
In the models which accrete significant amounts of gas, therefore,
the gas rapidly evolves to a condensed configuration, such as is likely to
lead to cooling of the gas, and greatly enhanced accretion by the individual
stars within the system.

We may make a crude analytical estimate of the accreted mass if we assume
that the potential acts as a ``cookie cutter,'' and that gas closer to the
center of the potential than $R_B$ is accreted during the passage of the
potential through the dense slab.  The amount of gas that would be
predicted to be accreted is then given by
\begin{equation}
M_p=\pi l R_B^2\rho_s,
\label{eq:mpred}
\end{equation}
where $\rho_s$ is the density of the gas in the slab.  The calculated values
of $M_p$ are shown in Table~\ref{tab:results}.

As can be seen from the table, the values of
$M_{acc}$ and $M_p$ are in fair agreement for some models, but not
others.  As indicated in \S2, the expression of $R_B$ in
equation~(\ref{eq:bondi}) is derived for a point-mass potential, and so is
only a good approximation for our models when $R_B$ significantly
exceeds $R_c$, a condition violated in Models~1 and 6.  For those models,
the Bondi radius computed assuming a point mass potential lies within
the core, where the Plummer potential actually flattens out.  The true
potential therefore never becomes as deep as required for substantial
accretion, and $M_{acc}\ll M_{p}$.
Because $u_\infty > C_s$ in Model 1, equation~(\ref{eq:mdotsum}) is more
appropriate to represent the gas accretion.  We could not, however,
include the fine grain structure of
the potential due to individual stars in our simulations.

In the case of Model~3, $R_B>R_t$.  Beyond $R_t$, the gravity of the
potential can have no direct effect upon the gas, and so
Equation~\ref{eq:mpred} represents an overestimate of the amount of gas
that is expected to be accreted.  Had we used $R_t$ instead
of $R_B$ in calculating the fiducial value of $M_p$, we would have found
$M_p=5.6\times10^7$~M$_\odot$, much closer to the value found for
$M_{acc}$.

For the remaining models, $M_{acc}>M_p$.  That discrepancy is likely to
be the result of the simplistic assumption made in deriving $M_p$,
that no hydrodynamic motion beyond $R_t$ results from the accretion.
That assumption is equivalent to assuming that $\tau_s>\tau_c$.
In fact, for most of the models, $\tau_s$ is at least comparable to,
and often substantially smaller than $\tau_c$, indicating that the gas
within the dense slab shall have time to undergo significant
adjustments to a new hydrostatic equilibrium in response to the
accretion.  As gas is accreted by the potential, therefore, the
resulting pressure gradient drives gas inward from beyond $R_t$,
increasing the effective radius from which gas may be accreted.  The
resulting enhancement in the accretion causes $M_{acc}>M_p$, even for
Model~6, for which $R_B<R_c$.

The results of these models indicate that the results of
Section~\ref{sec:analytic} represent reasonable estimates of the accretion
rates onto cluster potentials, and that the ratio
$\left(C_s^2+u_\infty^2\right)/\left(GM_d/R\right)$ is crucial to
determining the importance of accretion by the global potential.
The greatest differences between the analytic and numerical results occur at
low values of $\dot M$, where the accretion is insignificant.

\section{Astrophysical Applications}
\label{sec:applied}

We now consider three scenarios in hich accretion may affect the evolution of
stellar clusters.  These include potential affects upon the metallicity 
dispersion and white dwarf cooling sequences of globular clusters, the stellar
populations of clusters in the Galactic disk, and the rejuvenation of old
clusters near the Galactic center.

\subsection{Metallicity Dispersion and White Dwarf
Sequence in Globular Clusters}

Typical globular clusters in the Galactic halo have relatively high
orbital inclination and eccentricity.  In contrast, the interstellar
medium consists of either warm ($10^2$~K) atomic or cold (10~K) molecular gas
on nearly coplanar, circular Galactic orbits.  The relative velocity of
the halo clusters to the interstellar medium is in the range of
100~km~s$^{-1}$. For these clusters, the collective effect of their potential
is clearly insignificant, and cluster stars accrete as individual entities.

Scaling with an average number density of $10^2$~cm$^{-3}$ for atomic
clouds, we find from equation~(\ref{eq:mdots}) that $\dot M_\ast \sim 10^9 $g
s$^{-1}$.  Over the age of the Galaxy, the total amount of mass
accreted is $\sim 10^{-6} M_\odot$.  Even for the most metal-deficient
([Fe/H] $\simeq -4 $) stars known, the acquisition of this amount of
interstellar gas, with a solar metallicity, would modify the metallicity of
the star by less than 1\%.  Variations in Lithium abundance has been
observed among subgiants in halo globular clusters (Deliyannis, Boesgaard,
\& King 1995; Castilho et al. 2000).
Main sequence evolution almost certainly has depleted all the Lithium
in the envelopes of globular clusters' main sequence stars and the
above estimate for the accretion rate is too small to account for the
observed data. But, Lithium is produced in the core through hot bottom
burning and dredged up to the surface of stars on the asymptotic and red
giant branches (Cameron \& Fowler 1971; Sackmann, Smith, \& Despain 1974;
Sackmann \& Boothroyd 1992).
The consumption of a Jupiter-mass gas planet has also been suggested
as a mechanism to replenish the surface Lithium content (Laws \&
Gonzalez 2001; Sandquist et al. 2002),
though the depletion time scale for Lithium is relatively short.

For white dwarf stars, accretion at the above rate would lead to the
release of energy at a rate $L_a \sim 10^{-7} L_\odot$, which is much
smaller than that expected from the cooling of the oldest Galactic
white dwarfs (Hansen 1999).  Thus the accretion from the interstellar
medium needs not be taken into account in the determination of the
white dwarf cooling sequence.

In order, therefore, for accretion to have a significant effect upon either
the metallicity of cluster stars, or upon the the
color and luminosity of white dwarfs, $u_\infty$ must be reduced
to $\sim 10$ km s$^{-1}$, in which limit the effects of collective
accretion by the overall cluster potential would become important.

\subsection{Sedimentation of Clusters into Galactic Disks}

Although $u_\infty$ for typical halo clusters is too large for them to
accrete any significant amount of matter, their orbital properties may
evolve.
As halo clusters pass through the Galactic disk with a speed
$u_g$, the wake of disk stars induce a drag force (dynamical friction),
given by
\begin{equation}
F_d = 4 C_d \pi G^2 M_d^2 \rho_d u_g ^ {-2}
\end{equation}
where $\rho_d$ is the stellar density in the galactic disk, the drag
coefficient,
\begin{equation}
C_d = {\rm ln} \Lambda \left[{\rm erf}\left(X\right)-{2X\over{\pi^1/2}}
e^{-X^2}\right],
\end{equation}
where $\Lambda = b_{\rm max}/ R_c$, $X=u_g/\left(\sqrt{2}\sigma\right)$,
and $b_{\rm max}$ is the maximum effective impact parameter, which can be
taken to be the density scale height $H_d$ of the Galactic disk
(Chandrasekhar 1943; Binney \& Tremaine 1987).
Because gas and stars are concentrated near the midplane, the rate of
dissipation of the cluster's orbital energy, averaged over its Galactic
orbital period, $P_g$, is
\begin{equation}
\dot E = 2 \left({ F_d \over M_d} \right) \left( {H_g \over u_z} \right)
\left( {u_g \over P_g} \right)
\end{equation}
where $E= u_g^2/2$ is the kinetic energy relative to the disk and
$u_z$ is the velocity normal to the galactic disk.  Clusters which
cross the Galactic disk with large velocities would sediment into low
inclination and low eccentricity orbits on a time scale $u_g ^2 / 4 \dot
E$ (Artymowicz, Lin \& Wampler 1993). Over a time interval, $\Delta T$,
the critical condition for clusters to attain small $u_g$ and $u_z$,
and therefore $u_\infty$ is
\begin{equation}
\left({u_g \over V_c} \right)^3 {u_z \over V_c}
< 32 C_d \pi G^2 {M_d \Sigma_d \over
M_g(a)^2} { \Delta T  \over P_g}
\label{eq:sedi}
\end{equation}
where the Galactic circular velocity $V_c =\left[GM_g (a)/a\right]^{1/2}$
at a radius $a$ is determined by the Galactic mass $M_g (a)$ contained
within it, and $\Sigma_d = \rho_d H_d$ is the surface density of the
disk stars. For $V_c = 220$ km s$^{-1}$ at $a=10 $ kpc,
equation~(\ref{eq:sedi}) implies that clusters with $u_g \sim u_z < 0.3 V_c$
or $u_g = V_c$ and $u_z < 10^{-2} V_c$ would sediment into the disk
within the life span of the Galaxy.  Because $\Delta T$ is a decreasing
function of $u_z$ and $u_g$, the inclination and eccentricity of the
clusters' Galactic orbits quickly vanish, leading to a small $u_\infty$.

Based on the present phase-space distribution of globular clusters,
several authors (Fall \& Rees 1977; Keenan 1979; Ostriker \& Gnedin
1997; Vesperini, 1998; van den Bosch et al. 1999; Hideaki \& Yoshiaki 2003)
have suggested that dynamical friction may have caused clusters with
small initial orbital radii to undergo significant orbital evolution.  There
also exist clusters in a thick disk population, which have smaller values of
$u_\infty$ than typical halo clusters.

For clusters that are not on orbits highly inclined to the disk, collective
accretion may become a significant factor in their evolution.  As
$u_\infty$ is reduced to $\sim 10$ km s$^{-1}$, $\dot M_\ast$
approaches to $10^{13} $g s$^{-1}$.  More importantly, $u_\infty < (G
M_d / R_c)^{1/2}$, and $\dot M_c$ approaches $10^{25}$g s$^{-1}$.  The
gas density $\rho_c$ within the clusters' potential would increase
until
\begin{equation}
\dot M_\ast \sim (M_d/M_\ast) \dot M_c \sim 10^{-7}
(M_\ast/M_\odot)^2 M_\odot {\ \rm yr}^{-1}.
\end{equation}
Because clusters have nearly isothermal stellar distribution function,
$u_\infty$ relative to the gas in the cluster is comparable to
$\sigma$, and $\rho_c \sim (M_d/M_\ast) \rho_g \sim 10^{-16}$~g~cm$^{-1}$.
Gas within these clusters would achieve high densities.

Within this environment, the cluster stars would accrete gas and gain
significant mass.  Within one Galactic orbit, the observed metallicity of
the stars could attain the solar value present in the
interstellar medium.  Gas accumulating within the core of the cluster
potential may also trigger new star formation.  With high temperatures
and densities, the initial mass function may be heavily skewed toward
the high-mass range (McKee \& Tan 2003).  As star formation proceeds,
negative feedback from processes such as photoionization, stellar
winds, and supernovae can modify the gas concentration, and terminate
star formation \citep{DLM03}.  These negative feedback effects cannot,
however,
prevent re-accumulation of gas around the sedimented clusters after
the newly formed massive stars evolve off the main sequence.  The
distinguishing properties of these sedimented clusters from open
clusters are their centralized cores and rich population of white
dwarfs, despite their gas rich environment and the pre-main-sequence
color magnitude diagram.

There are no stellar clusters observed in the Galactic disk which bear
these anticipated properties, as would be expected given the relatively
small region of phase space occupied by clusters whose orbits would be
heavily modified by dynamical friction.  At early epochs, however, many
more such clusters would be expected, as a natural
extrapolation of the cold dark matter scenario in which merger events
and the subsequent dynamical friction would lead to the dynamical
evolution of not only globular clusters but also the cores of dwarf
galaxies (Moore et al. 1999; Klypin et al. 1999; Metcalf 2002;
Metcalf \& Zhao 2002; Murray et al. 2003).  Star formation via accretion
onto stellar and dark matter potentials may therefore have played an
important role in the early evolution of galaxies.

Accretion at lower levels than the maximum estimated above would still
affect the white dwarf cooling sequences of clusters.  Clusters with orbits
on small radii, and close to the galactic plane might therefore be expected
to have white dwarf cooling sequences brighter than expected for their ages.

\subsection{Accretion Onto Remnant Cores of Globular Clusters and Nucleated
Dwarf Galaxies}

Near the Galactic center, several young massive star clusters have been
recently discovered (Nagata et al. 1990, 1995; Okuda et al. 1990;
Figer et al. 1999).  Besides their age, these cluster are unusual in that
1) the initial mass function of stars within them is highly skewed to
the high mass range (Figer et al. 2002; Stolte et al. 2002); and 2)
they have short dynamical life expectancy (Kim, Morris, \& Lee 1999;
Kim et al. 2000; Portegies Zwart McMillan, \& Gerhard 2003; MacMillan
\& Portegies Zwart 2003).  One cluster, the Arches Cluster, contains
up to 7$\times10^4$~M$_\odot$ within a radius of 0.23~pc, and has a
velocity dispersion of up to 22~km~s$^{-1}$ (Figer et al. 2002).  In
addition, the Quintuplet Cluster, the cluster around the Galactic center
itself, and several other candidate clusters (Law \& Yusef-Zadeh 2004) contain
x-ray point sources.  The presence of clusters of young, short-lived stars
near the Galactic center is remarkable, given that it is an environment
that is extremely hostile to the formation of star clusters.

Assuming that the above objects are all young and short-lived clusters,
and that there is nothing unusual about the current epoch, then
more than $10^4$ such clusters would be expected to have existed over
in the lifetime of
the Galaxy.  The inferred heavy element generation by the prolific
production of massive stars would enrich the Galactic center region
to well above the present level, unless gas is effectively removed from
that region (cf Figer et al. 2004).

Based on the results of the current work, we suggest an alternative
scenario for
the origin of these extraordinary clusters.  Under the influence of
dynamical friction, globular clusters undergo orbital decay towards
the Galactic center. As discussed above, dynamical friction is not 
important to the evolution of most clusters.  It is computed, however, to
have a significant effect for clusters within a few kpc of the Galactic
center (e.g. Fall \& Rees 1977).  The evolution of such clusters, and
accretion of gas by them, may explain the existence of young clusters of
massive stars near the Galactic center.

During the course of the clusters' orbital
decay, the Galactic tidal field gradually increases and their tidal
radius gradually shrinks.  Stars close to or outside of the tidal radius
become detached while those near the dense core are retained (Oh \& Lin
1992; Oh, Lin, \& Aarseth 1995).  Stellar loss also shortens the two-body
relaxation time scale, enhances the central concentration, and shortens
the tidal disruption time scale (Kim \& Morris 2003).  With a
core density comparable to that of post-collapse clusters M15
(Guhathakurta et al. 1996) and M30 (Yanny et al. 1994),
systems with more than $10^4$ stars can persist and be tidally
preserved near the present location of Galactic center clusters.  Less
centrally dense clusters such as M13 (Cohen et al. 1997) are
likely to be disrupted by the Galactic tide outside 30 pc.

We suggest that Arches and Quintuplet clusters are the rejuvenated cores of
old globular clusters, which have retained their
integrity as their orbits decay to the vicinity of the Galactic center.  Of
those, however, only a fraction have velocities relative to the atomic and
molecular clouds that are sufficiently small for collective accretion to
become significant.
The velocity dispersion of the stars and gas clouds within 30 pc from the
Galactic center is $\sigma_G \sim 50$km s$^{-1}$ (Genzel, Hollenbach
\& Townes 1994).
The central velocity dispersion of pre-collapse cores is typically
$\sigma_\ast \sim 10$km.  The fraction of clusters with $C_\infty^2 +
u_\infty^2 < \sigma_\ast^2$ (cf the condition set in equation~\ref{eq:bcond})
is $\sim$ erf($\sigma_\ast^2/ \sigma_G^2) \sim
\sigma_\ast^2/ \sigma_G^2$, which is on the order of a few percent.

For clusters having $M_d \sim 10^4 M_\odot$ and sufficiently small speed
relative to the ISM, 
we find from equation~({\ref{eq:mdotc1}) that the gas accretion rate
onto the cluster $\dot M_c > 10^{23}$~gm~s$^{-1}$ in regions where the
density of
the ambient gas near the Galactic center is $>10^4$ cm$^{-3}$
(Launhardt, Zylka, \& Mezger 2002). The accumulation of gas near the center
of the cluster potential enables solar-type stars to accrete at a rate
$\sim 10^{-6} M_\odot$ yr$^{-1}$.  Because the accretion rate is proportional
to $M_\ast^2$, the growth rate of the stars accelerates rapidly as they accrete
mass.  That rapid acceleratoin allows stars with masses of 20~M$_\odot$ to be
built up within $\sim$1~Myr, by accretion onto stars with initial masses
comparable to that of the Sun, a timescale shorter than the lifetimes of the
massive stars formed via accretoin.  With such a small velocity relative to
the dense gas, the clusters do not
drift more than a few pc from the orbits of any clouds within 1~Myr, which
would allow the low-mass mature stars to be rejuvenated into young,
massive stars.

The main advantage of this scenario is that it does not require the
rapid formation of clusters with unique initial mass functions. In
comparison with star forming regions in the Solar neighborhood, the
Galactic center is a challenging environment for efficient star
formation.  Our scenario naturally bypasses the stringent conditions for
forming the progenitor clouds in regions where the external tidal
effect is strong and heating is intense.  This scenario requires a combination
of clusters surviving tidal destruction as their orbits decay, and
encountering dense gas with sufficiently low relative velocity to lead to
significant accretion.  Such a combination of events is certainly relatively
unlikely.  However, in the cold dark matter scenario, the initial population
of clusters and dwarf galaxy cores near the center of the Galaxy was much
larger than today, such that the population of faint, old clusters 
near the Galactic center would be significantly larger than predicted from
the current population of globular clusters.
A fairly unlikely sequence of events is therefore needed if
we are to avoid having many more systems such as the Arches and Quintuplet
clusters today.

Additionally,
the centers of dense clusters are populated with both white dwarfs and
neutron stars (Yanny et al. 1997).  Accretion onto these dense
stellar cores at the inferred rate can lead to the onset of luminous
x-ray sources.  Such a process may account for the distinct point
sources of X-ray emission found by Chandra in these clusters (Laws \&
Yusef-Zadeh 2004).

\section{Summary and Discussion}
\label{sec:discussion}
In this paper, we examine in detail the accretion of ambient gas by a
stellar system.  Although the gravitational potential of the
individual stars are point mass, that of the cluster is softer.  We
showed that if either the relative speed of the cluster or the sound speed of
the ambient gas is large compared with the velocity dispersion of the
core, the accretion by the cluster is inefficient.  Individual stars
accrete gas as though they move through the interstellar medium
independently.  But if the relative speed of the cluster and the sound
speed of the ambient gas is less than that of the internal velocity
dispersion, gas is accreted into the cluster potential collectively
and rapidly. Accretion by individual stars is then enhanced greatly
relative to their rate of accretion directly from the ambient gas.

We suggest that this process may be important in inducing chemical
inhomogeneity and modifying the white dwarf cooling sequence some
globular clusters whose orbits happen to lie close to the Galactic plane.
We also speculate that this process may have caused some
tidally stripped cores of globular clusters near the Galactic center
to accrete gas and to rejuvenate their member stars, resulting in the
formation of clusters of young, hot stars, such as are seen near the center
of the Galaxy.

\begin{acknowledgements}
This work was performed under the auspices of the U.S. Department of
Energy by University of California, Lawrence Livermore National
Laboratory under Contract W-7405-Eng-48.  This work is partially
supported by NASA through an astrophysical theory grant NAG5-12151.
\end{acknowledgements}

\clearpage
\begin{deluxetable}{cccccccc}
\tablewidth{0pt}
\tablecaption{Models \label{tab:results}}
\tablehead{
\colhead{Model} &
\colhead{$V$} &
\colhead{$c_s$} &
\colhead{$R_B$} &
\colhead{$\tau_c$} &
\colhead{$\tau_s$} &
\colhead{$M_p$} &
\colhead{$M_{acc}$} \\
\colhead{} &
\colhead{km~s$^{-1}$} &
\colhead{km~s$^{-1}$} &
\colhead{pc} &
\colhead{Myr} &
\colhead{Myr} &
\colhead{M$_\odot$} &
\colhead{M$_\odot$}
}
\startdata
1 & 30 &  1 &   1.7 &   3.8 &  1.65 & 2.4$\times10^4$ & 3.4$\times10^1$ \\
2 & 10 &  1 &  14.8 &  11.4 & 14.6  & 2.0$\times10^6$ & 6.0$\times10^6$ \\
3 &  1 &  1 & 740   & 114   &730    & 4.9$\times10^9$ & 2.2$\times10^7$ \\
4 & 10 & 10 &   7.4 &  11.4 &  0.73 & 4.9$\times10^3$ & 2.1$\times10^4$ \\
5 &  1 & 10 &  14.8 & 114   &  1.46 & 2.0$\times10^4$ & 3.2$\times10^4$ \\
6 &  1 & 30 &   1.6 & 114   &  0.054& 2.4$\times10^1$ & 3.4$\times10^3$
\enddata
\end{deluxetable}

\clearpage
\begin{figure}
\plotone{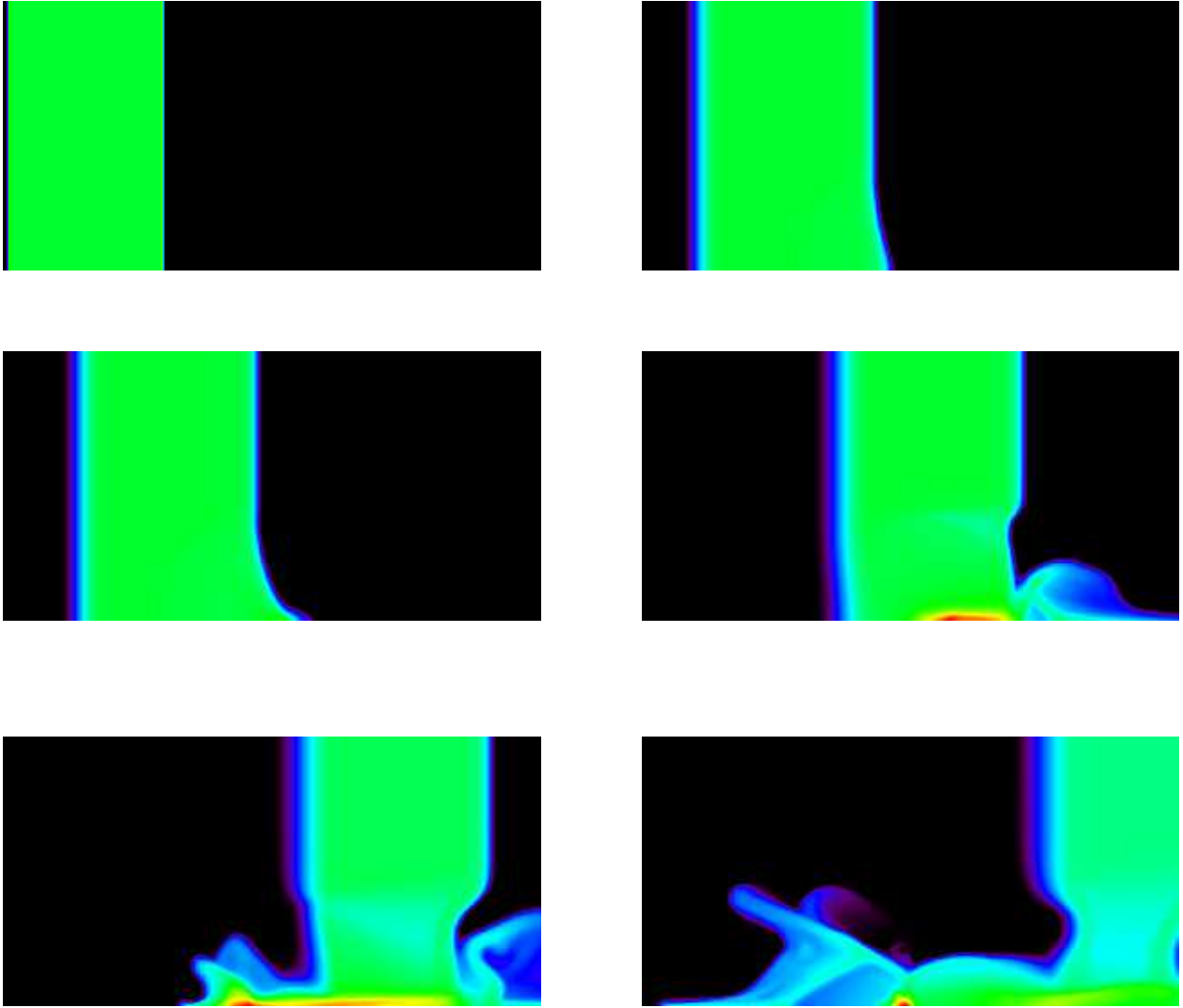}
\caption{The density evolution of Model~2.  The model is shown at the
times 0, 4.5, 6.0, 7.5, 22.5, and 30~Myr.  The dense cloud, modelled as a
simple slab of gas, sweeps from left to right across the accreting potential,
centered in the model.  Because of the relatively deep potential, and slow
speed relative to the cloud, the potential has a significant effect upon the
dense gas.  Some of the gas is initially pulled ahead of the rest of the cloud
into the potential, and much of the gas within the tidal radius is accreted
as the cloud passes by the potential, where it forms the small, dense core
seen in the final frame.}
\label{fig:model2rho}
\end{figure}

\end{document}